\documentstyle[twocolumn,aps,epsfig]{revtex}

\begin{document}
\draft
%\preprint{HEP/123-qed}
\twocolumn[\hsize\textwidth\columnwidth\hsize\csname@twocolumnfalse\endcsname

\title{Intersite coupling effects in a Kondo lattice}
\author{S. Nakatsuji$^{1}$, S. Yeo$^1$, L. 
Balicas$^{1}$, Z. Fisk$^{1}$, P. Schlottmann$^{2}$, P.G. 
Pagliuso$^{3}$, N.O. Moreno$^{3}$, J.L. Sarrao$^{3}$, \\ and J.D. 
Thompson$^{3}$}
\address{$^1$ National High Magnetic Field Laboratory (NHMFL), 
Florida State 
University, Tallahassee, Florida 32310 \\
$^2$ Department of Physics, Florida State University, Tallahassee, 
Florida 32306 \\
$^3$ Los Alamos National Laboratory, Los Alamos, New Mexico 87545\\}

\date{May, 23, 2002}
\maketitle

\begin{abstract}
The La dilution of the Kondo lattice CeCoIn$_5$ is studied. The
scaling laws found for the magnetic susceptibility and the specific 
heat reveal two well-separated energy scales, corresponding to the 
single impurity Kondo temperature $T_{\rm K}$ and an intersite 
spin-liquid temperature 
$T^*$. The Ce-dilute alloy has the expected Fermi liquid ground 
state, while the specific heat and resistivity in the dense Kondo 
regime exhibit non-Fermi-liquid behavior, which scales with $T^*$.  
These observations indicate that the screening of the magnetic 
moments in the lattice involves antiferromagnetic intersite 
correlations with a larger energy scale in comparison with the Kondo 
impurity case.
\end{abstract}
]

\narrowtext
The nonuniversal behavior of heavy fermion compounds has challenged 
experimental and theoretical physicists over many years. A wide range
of low-temperature phenomena, such as strongly enhanced 
paramagnetism, 
magnetic/quadrupolar long-range order, Kondo 
insulators, unconventional superconductivity, and non-Fermi-liquid 
behavior, have been reported. This variety of phenomena is believed 
to arise through the competition and interplay of the Kondo effect, 
band structures and intersite correlations. Both, the Kondo-esque 
resonance at the Fermi level \cite{Hewson} and the 
Ruderman-Kittel-Kasuya-Yosida (RKKY)
interaction in a Kondo lattice, are the consequence of the large 
Coulomb repulsion
in the partially filled 4$f$(5$f$)-shell in conjunction with the 
hybridization of the $f$-electrons with the conduction state.

In the single impurity limit, the low energy transport and the 
thermodynamic properties scale with a single parameter, the Kondo 
temperature $T_{\rm K}$ \cite{Hewson}. As $T$ is decreased below 
$T_{\rm K}$, the moments which are localized at high temperatures 
are screened into local singlets by the conduction electrons, 
giving rise to a local Fermi liquid state. The agreement between 
experiment and theory is excellent for many dilute systems 
\cite{schl}.

For Kondo lattice systems, it has been a long-standing issue 
how the intersite interactions affect the local character of the 
Kondo screening.
If the singlet formation in the lattice involves the neighboring 
localized moments, 
it is expected that the spin screening is accompanied by 
antiferromagnetic (AF) 
short-range correlations, which may modify $T_{\rm K}$ or yield a new 
energy scale. 
While a study 
of the Kondo lattice within the Gutzwiller approximation 
\cite{RiceUedaPRL} 
predicts such a ``lattice enhancement'' of $T_{\rm K}$, $1/N$ 
expansions \cite{1/Nexpansion} indicate local singlet formation with 
no change in the energy scale. The lattice enhancement 
has been confirmed by Monte Carlo, exact 
diagonalization, and density matrix renormalization-group 
\cite{1DKLMreview} for the 1 D Kondo lattice, but 
backscattering across the Fermi surface may affect this result. 
Monte Carlo studies for the 2D half-filled Kondo 
lattice yield different 
renormalizations of the spin and charge gaps \cite{Assaad}.

Experimentally, the strength of the intersite coupling can be varied
by substituting the magnetic Ce ions by their nonmagnetic analog
La. This approach continuously interpolates between the Kondo
lattice and the single impurity limit. Several systems have been 
studied, e.g. (Ce,La)Al$_2$, (Ce,La)B$_6$, and (Ce,La)Cu$_6$ 
\cite{Ladilution}, but it is hard to arrive at quantitative 
conclusions, since La substitution often changes both the 
crystalline electric field (CEF) parameters and Kondo coupling.

The ideal conditions of concentration independent $T_{\rm K}$ and 
CEF are met for (Ce,La)Pb$_3$ \cite{LinPRL}. It was reported that 
the specific heat and susceptibility scale with the Ce-concentration
for the entire alloying range and follow the predictions of the 
$S=1/2$ Kondo impurity model with $T_{\rm K} \simeq 3$ K 
\cite{LinPRL}. 
No signs of intersite couplings were found, except in the 
concentrated limit, where they eventually lead to the AF ordering
at $T_{\rm N} \simeq 1$ K for CePb$_3$. 

In this Letter we present experimental evidence for three 
well-separated 
energy scales in Ce$_{1-x}$La$_x$CoIn$_5$: $T_{\rm K}$, the energy 
scale for intersite 
interactions $T^*$, and the level splitting due to CEF.
CeCoIn$_5$ is a heavy fermion unconventional superconductor with a 
high transition 
temperature $T_{\rm SC} =$ 2.3 K 
\cite{Petrovic,ShishidoJPSJ,Sidorov}. With La dilution, 
Ce$_{1-x}$La$_x$CoIn$_5$ has a paramagnetic ground state with Ce
keeping essentially the same valence and CEF splitting 
through the entire alloying range. We found scaling functions with 
the energy scale $T^*$ 
for the deviations of the specific heat and the susceptibility from 
the single impurity behavior. 
This scale $T^*$ is attributed to intersite interactions. 
The specific heat in the dense Kondo regime exhibits non-Fermi-liquid 
behavior  
with the characteristic temperature $T^*$, 
while for dilute alloys it follows that of the $S=1/2$ Kondo impurity 
model.  
The deviation of the susceptibility from the single ion behavior 
shows a strong decrease below $T^*$, suggesting the formation of a 
spin-liquid. 
The scaling procedure reveals that $T^*$ systematically increases 
from the 
single-ion $T_{\rm K}$ with increasing Ce concentration.
This indicates that the screening process of the localized moments in 
the lattice 
has higher energy scale, involving AF intersite correlations.

\begin{figure}[htb]
\begin{center}
\epsfig{file=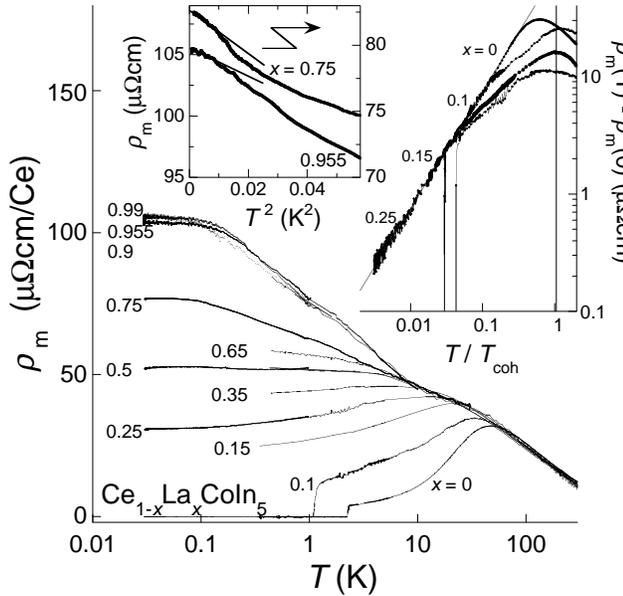, angle = 0, width=8.5cm}
\caption{Magnetic in-plane resistivity $\rho_{\rm m}$ for various 
$x$ of Ce$_{1-x}$La$_x$CoIn$_5$. Right inset: The log-log plot 
for the inelastic part of $\rho_{\rm m}$ vs. $T/T_{\rm coh}$; 
the solid line is the $T$ linear fit and the vertical lines mark 
the onset of superconductivity. Left inset: $\rho_{\rm m}$ vs. 
$T^2$ for the incoherent regime.}
\end{center}
\end{figure} 

Single crystals were prepared using an In self-flux method 
\cite{Petrovic,structure}. Before growing the crystals, Ce and La 
metals are premixed by arcmelting at least 10 times to 
get a homogeneous mixture. X-ray powder diffraction patterns show a 
single phase tetragonal HoCoGa$_5$ structure, which can be viewed 
as alternating layers of (Ce,La)In$_3$ with layers of CoIn$_2$ 
along the $c$ axis. The resistivity was measured by 
standard four-probe dc and ac techniques down to 25 mK using the 
low temperature facilities at the NHMFL. 
The magnetization was measured down to 1.8 K with a Quantum Design 
MPMS SQUID magnetometer and the specific heat 
($C_P$) by a thermal relaxation method using a Quantum Design PPMS.

Figure 1 shows the magnetic part of the in-plane resistivity 
$\rho_{\rm m}$ defined as the difference in $\rho(T)$ of 
(Ce,La)CoIn$_5$ 
and LaCoIn$_5$ (the nonmagnetic, isostructural, analog of CeCoIn$_5$) 
divided by the Ce concentration, $1-x$. Above 50 K, the $\rho_{\rm 
m}$ 
data for all $x$ collapse onto a single curve with the $-$log $T$ 
dependence characteristic of the Kondo effect. Hence, the high $T$ 
Kondo temperature, $T_{\rm K}^{\rm h}$, is essentially independent 
of $x$. In the Ce rich region, $\rho_{\rm m}$ shows coherent metallic 
behavior after rising through a peak that defines the coherence 
temperature $T_{\rm coh}$. The La dilution reduces $T_{\rm coh}$,
which tends to zero near $x$ = 0.5, where $\rho_{\rm m}$ saturates 
below 500 mK. With further La dilution, $\rho_{\rm m}$ evolves 
toward single ion behavior with a log $T$ dependence down to 500 mK
that suggests a small $T_{\rm K}$ of about 1-2 K. Hence, $x=0.5$ 
roughly separates the coherent and single ion regimes. It is 
interesting to note that 50 \% is close to the percolation limit of 
41 \% of a 2D square lattice (consistent with the 2D nature 
of Ce-Ce nearest neighbor network in the CeIn$_3$ layers).

\begin{figure}[tb]
\begin{center}
\epsfig{file=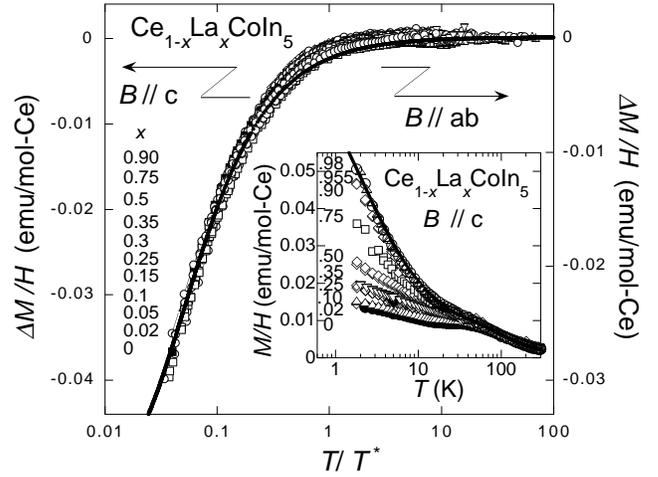, angle=0, width=8.5cm}
\caption{Excess susceptibility $\Delta\chi(T) \equiv \chi(T) - 
\chi_0(T)$ ($\chi_0(T)$ refers to the single impurity limit at
$x$ = 0.955), scaled with $T/T^*$. The solid line corresponds to 
an activation type fit. Inset: the $c$-axis component of $M/H$ 
versus log$T$.}
\end{center}
\end{figure}

The inset of Fig. 2 shows the $c$-axis component of the 
susceptibility 
$\chi_c(T)$ per mole Ce vs. log$T$. As for $\rho_{\rm m}$, the data
above 80 K collapse onto the same curve for all concentrations, 
again indicating that $T_{\rm K}^{\rm h}$ is independent of $x$. 
The high $T$ data follows the modified Curie-Weiss law $\chi(T)=C/(T 
+ \sqrt2 T_{\rm K}^{\rm h})$ \cite{Hewson} with $T_{\rm K}^{\rm h} 
\approx 35$ K and an effective paramagnetic moment of 2.5 $\mu_{\rm 
B}$. In contrast, at low $T$, $\chi(T)$ increases with La dilution, 
reaching the single-impurity limit for $x \geq 0.95$. Similar 
systematics are also observed for the ab-plane component, 
$\chi_{ab}(T)$. 

We performed a CEF fit to the ab-plane 
and c-axis components of the susceptibility for the dilute Ce regime
(see solid line in the inset of Fig. 2). The ground state is found 
to be a $\Gamma_7^{(1)}$ doublet $(-0.39 |\pm5/2\rangle + 0.92 
|\mp3/2\rangle$, including a low-temperature Weiss 
temperature of 2 K) with a fairly large gap to the excited 
$\Gamma_7^{(2)} (0.92 |\pm 5/2\rangle + 0.39 |\mp 3/2\rangle,\Delta_1 =$ 148.0 K) 
and $\Gamma_6 (|\pm1/2\rangle, \Delta_2 =$ 196.5 K) doublets. 
This scheme also well fits the high $T$ susceptibility above 80 K for 
all $x$ for both the ab-plane and c-axis components, which suggests 
that the CEF scheme is essentially independent of $x$. 
In order to confirm the CEF scheme, 
we measured the specific heat up to 200 K for $x$ = 0, 0.25, 0.50, 
0.75, 1.00. The magnetic 
contribution to the specific heat $C_{\rm m}$ is estimated as 
$C_P$ of Ce$_{1-x}$La$_{x}$CoIn$_5$ minus $C_P$ of LaCoIn$_5$ 
divided by the Ce concentration. 
For all $x$, $C_{\rm m}$ shows a broad peak around 100 K, 
consistent with the Schottky type fit based on the above CEF scheme 
(see Fig.3 right inset).
These data also imply that the CEF scheme is roughly independent of 
$x$.
There are differing CEF fits for CeCoIn$_5$ in the literature. Our 
results agree best with the scheme of Ref. \cite{ShishidoJPSJ} and 
less well
with those in Refs. \cite{CEF}. We argue that these discrepancies 
arise from the way intersite correlations are taken into account. 

\begin{figure}[htb]
\begin{center}
\epsfig{file=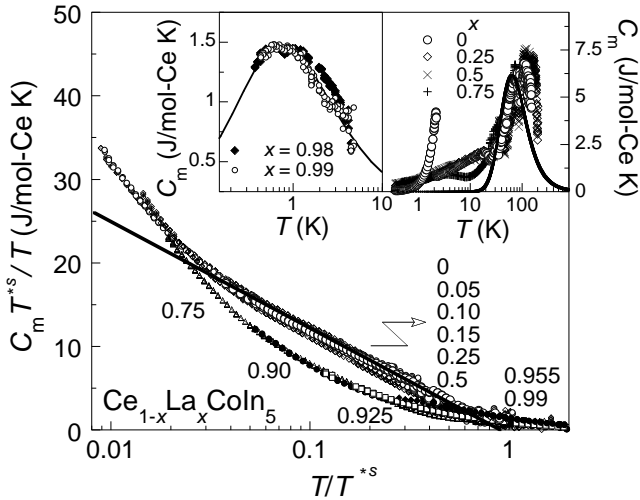, angle=0, width=8.6cm}
\caption{$C_{\rm m}/T$ times $T^{*s}$ vs. $T/T^{*s}$. 
The solid line represents the $-$ln$T$ fit. 
Left inset: $T$ dependence of $C_{\rm m}$ for $x$ = 0.98 and 0.99. 
The solid curve is the fit to the $S = 1/2$ Kondo impurity limit with 
$T_{\rm K} = 1.7$ K. 
Right inset: $T$ dependence of $C_{\rm m}$. The solid curve is the 
fit based on our CEF scheme.}
\end{center}
\end{figure}

Given the essential same CEF scheme for all $x$, 
we interpret the change in $\chi(T)$ with La dilution at low $T$ as 
an intersite coupling 
effect. To quantify this contribution, we subtract the 
single impurity CEF $\chi(T)$ from $\chi(T)$ for each $x$ to get the 
``excess susceptibility'' 
$\Delta\chi(T)$, which shows a strong decrease of $\chi$ at a 
concentration dependent characteristic temperature $T^*$ (e.g.,  
around 50 K for CeCoIn$_5$). Interestingly, we found a scaling 
law for both $ab$ and $c$ components of $\Delta \chi(T)$ vs. 
$T/T^*(x)$ as shown 
in Fig. 2. Basically the same systematic change
of $T^*$ with $x$ is found for both components of 
$\chi$, $T^{*ab}$ and $T^{*c}$, as seen in Fig. 4. 

For Kondo systems $C_{\rm m}/T$ is given by 1/$T^{*s}$ ($T^{*s}$ 
is a characteristic energy) times a scaling function of $T/T^{*s}$ 
\cite{Hewson}. Actually, we have found two different scaling 
functions in Ce$_{1-x}$La$_{x}$CoIn$_5$. In the coherent 
region ($x < 0.5$), for the temperature range between 0.04$T^{*s}$ 
and 0.6$T^{*s}$, $C_{\rm m}/T$ fits well to the expression 
$-5.4/T^{*s}$ln$(T/T^{*s })$ (J/mol-Ce K$^2$) as displayed 
by the solid line in 
Fig. 3. Considering 5.4 J/mol-Ce K is close to $R\ln2$, this 
expression suggests that 
$T^{*s}$ gives the temperature scale to recover the ground 
doublet entropy.
In the incoherent regime ($x > 0.5$), 
all the curves scale to another function as shown in Fig. 3. 
Furthermore, 
the $C_{\rm m}/T$ data for $x$ = 0.98 and 0.99 agrees well with the 
exact results for the $S = 1/2$ 
single-ion Kondo model with the single parameter $T_{\rm K} = 1.7$ 
K, which defines the $T^{*s}$ at the single impurity limit 
\cite{ExactSolution} (see left inset of Fig.3). 

Generally, $T_{\rm K}$ is related to the high-temperature 
$T_{\rm K}^{\rm h}$ and the CEF splittings via $k_{\rm B}T_{\rm K} =
(k_{\rm B}T_{\rm K}^{\rm h})^3/\Delta_1\Delta_2$ 
\cite{YamadaHanzawaYoshida}.
From the susceptibility we have $T_{\rm K}^{\rm h} \approx 35$ K, 
$\Delta_1 = 148$ K, and $\Delta_2 = 196.5$ K, such that the 
expression 
yields $T_{\rm K} \approx 1.5$ K, consistent with the $T^{*s}$ 
and the 
low $T$ Weiss temperature in the single impurity limit.

\begin{figure}[tb]
\begin{center}
\epsfig{file=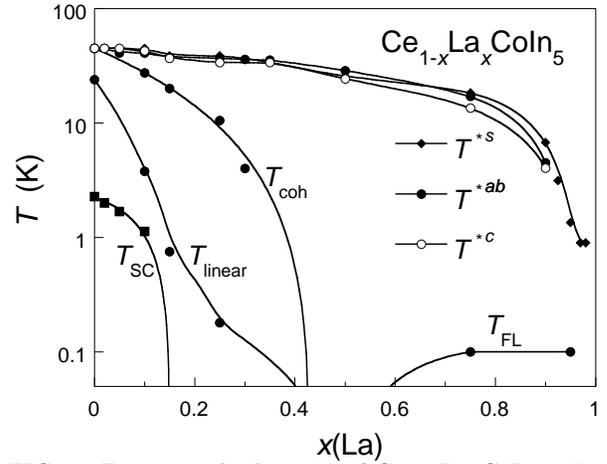, angle=0, width=8cm}
\caption[Phase diagram of Ce$_{1-x}$La$_x$CoIn$_5$]
{Energy scale diagram of Ce$_{1-x}$La$_x$CoIn$_5$. 
$T_{\rm SC}$ is the superconducting transition temperature determined 
by the onset of the jump observed in $C_{\rm P}(T)$.
$T_{\rm linear}$ and $T_{\rm FL}$ are the temperatures below which 
the resistivity starts to show $T$ linear and $T^2$ dependence.
$T^{*ab}$, $T^{*c}$ and $T^{*s}$ are the characteristic 
temperatures of the ab-plane and c-axis susceptibilities, and 
the specific heat, respectively.}
\label{PhaseDiagram}
\end{center}
\end{figure}

With the scaling procedures above, we obtain 
a systematic La concentration dependence of $T^{*s}$ as shown in 
Fig. 4. 
There are three points to note. 
First, the characteristic temperatures $T^{*s}$, $T^{*ab}$ and 
$T^{*c}$ are essentially identical, indicating that there is only 
one energy scale besides $T_{\rm K}$. Second, in the $x \to 1$ limit 
this 
energy scale agrees with $T_{\rm K}$, showing that 
the scale $T^*$ originates from the single ion $T_{\rm K}$.
Third, in the dense Kondo limit CeCoIn$_5$, $T^*$ coincides 
with the $T_{\rm coh}$ of the resistivity. 

Because $T_{\rm K}^{\rm h}$ is constant throughout the alloy series, 
and the high $T$ 
$\chi$ and $C_{\rm P}$ data point to no changes in the CEF 
parameters, the single ion 
$T_{\rm K}$ should also be constant. {\it The systematic increase 
in $T^*$ must then arise from the intersite coupling}. Hence,
this system has three well-separated energy scales: $T_{\rm K} 
\approx 1-2$ K for the single ion Kondo screening, $T^* \approx 
40-50$ K for Kondo lattice effects and the CEF excitations $\Delta 
\approx 200$ K. The fact that these energy scales are very different 
is a fortuitous situation that allows the present scaling analysis.

The different scaling of $C_{\rm m}/T$ in the coherent and 
incoherent regimes suggests that the nature of the ground state 
changes near $x$ = 0.5. A qualitative change was also found in the 
low-$T$ dependence of the resistivity. The right inset of Fig. 1 
shows the log-log plot of the inelastic part of the in-plane 
resistivity vs. $T/T_{\rm coh}$ in the coherent regime. All curves 
for $x \leq 0.25$ have a $T$-linear dependence below a characteristic 
temperature $T_{\rm linear}$. Note that all the $T$-linear curves 
have the same coefficient $A/T_{\rm coh}$, which reveals that the 
inelastic scattering rate scales with the $1/T_{\rm coh}$. Especially 
for CeCoIn$_5$, 
it scales with $1/T^*$ because $T_{\rm coh} = T^*$ at $x \to 0$ 
limit.   

In contrast, the left inset of Fig. 1 displays the $T^2$ dependence 
for $x = 0.75$ and 0.955 in the incoherent regime, which becomes 
evident below the Fermi liquid temperature $T_{\rm FL} \simeq 100$  
mK. 
In the single impurity limit, according to Nozi\`{e}res and Yamada 
\cite{NoziereYamada} the $T^2$ coefficient of the conductivity should 
be proportional 
to $T_{\rm K}^{-2}$. Using their formula, 
$T_{\rm K}$ is estimated to be 1.1 K for $x = 0.955$. This is 
consistent with the estimates above.

Non-Fermi-liquid (NFL) behavior is manifested in the $T$-linear 
resistivity and $-\ln(T/T^*)$ dependence of $C_{\rm m}/T$ 
(Figs. 1 and 3). NFL behavior has been observed in numerous other
Ce, Yb and U alloys and compounds \cite{Stewart} and is frequently
attributed to a quantum critical point (QCP) due to the proximity of 
an 
AF instability. 
Especially near a QCP of 2D antiferromagnetism, critical fluctuations 
are expected to lead to 
$\rho \propto T/ T^*$ and $C_{\rm m}/T \propto -\ln(T/T^*)$ 
\cite{Stewart,Moriya} as experimentally observed. In this case, $T^*$ 
gives the energy scale of the spin fluctuations 
\cite{Stewart,Moriya,Coleman}.

In addition, 
several experiments actually point to the 2D AF critical fluctuations.
De Haas-van Alphen measurements have shown that 
Ce$_{1-x}$La$_x$CoIn$_5$ has a strong 2D anisotropy of the 
Fermi surfaces reflecting its layered structure 
\cite{ShishidoJPSJ,Donavan}. 
The resistivity and specific heat under pressure 
\cite{ShishidoJPSJ,Sidorov}
and NMR $1/T_1$ measurements \cite{Kohori} suggest that CeCoIn$_5$ is 
close to the QCP of quasi-2D antiferromagnetism. Optical measurements 
of CeCoIn$_5$ have revealed 
the formation of an additional peak below 50 K within the 
hybridization 
gap \cite{Singley}, which is attributable to the quasi-particles 
interacting with AF fluctuations. Notably, their estimate of the 
fluctuation energy yields the same energy scale (8 meV) as $T^*$ for 
CeCoIn$_5$ (45 K). 

Moreover, the decrease of the excess susceptibility $\Delta \chi$ 
below $T^*$ is almost isotropic as shown in Fig. 2, 
and suggests the development of RVB-like singlets (short-range AF 
correlations of $f$ moments).
The $\Delta \chi(T)$ is well-described by the expression $\Delta 
\chi(0) [1-\exp(-0.05 T^*/T)]$ with $\Delta \chi_{ab}(0) \approx 
-0.038$ emu/mol-Ce and $\Delta \chi_c(0) \approx -0.052$ emu/mol-Ce. 
The reduced activation energy $\Delta = 0.05 T^*$ suggests slightly 
dispersive AF correlations. 
Interestingly, the $\Delta$ for CeCoIn$_5$ gives the same energy 
scale as the possible psuedo gap temperature $(\approx 3$ K) where 
the resistivity decrease was observed \cite{Sidorov}.

These results suggest that spin liquid formation with the gap 
$\Delta$ starts through the screening process, involving the 
intersite AF correlations with the energy scale $T^*$. 
Furthermore, the qualitative change in the $T$ dependence of both 
$C_{\rm m}/T$ and $\rho_{\rm m}$ around $x = 0.5$ 
indicates that a Ce concentration larger than 50 \% is the criterium 
not only for coherent transport but also for the intersite coupling 
leading to a short-range collective mode. This implies that the 
correlation length is of the order of the lattice constant $a$ and is 
rather short compared to what is usually expected for the RKKY interaction.

In summary, we have studied the effects of La dilution on CeCoIn$_5$. 
The scaling laws found in the susceptibility and specific heat 
reveal a systematic evolution of the characteristic intersite 
coupling energy $T^*$. Below this temperature, we found a Fermi 
liquid state in the incoherent impurity regime, while NFL behavior, 
which scales with $T^*$, has been observed in the coherent dense 
Kondo regime. The latter suggests that the Kondo screening in the 
lattice is no longer local but rather an intersite effect involving 
AF spin fluctuations. Given that the ratio of $T_{\rm SC}$ to the 
spin fluctuation energy $T^*$ for CeCoIn$_5$ ($\approx$ 5 \%) is of the 
same order as those for cuprates and other heavy fermion 
superconductors \cite{Moriya},
it is suggestive to associate $T^*$ with the energy scale leading to 
the relatively 
high $T_{\rm SC}$ of CeCoIn$_5$.

The authors acknowledge D. Hall, T. Murphy, and E. Palm for technical 
support and discussions, and G. Martin, K. Ueda, 
K. Yamada, and C.M. Varma for comments. This work 
was performed at the NHMFL, which 
is supported by NSF Cooperative Agreement No. DMR-9527035 and by 
the State of Florida. This work is also supported by NSF and DOE through grants 
Nos.DMR-9971348, DMR-0105431 and DE-FG02-98ER45797. S.N. has been supported by JSPS 
Research 
Fellowship.

\end{document}